\documentclass[12pt,english,aps,manuscript]{article}
\usepackage[T1]{fontenc}
\usepackage[latin1]{inputenc}
\usepackage{geometry}
\usepackage[active]{srcltx}
\usepackage{amsmath}
\usepackage{amssymb}
\usepackage[numbers]{natbib}

\makeatletter

\usepackage{geometry}

\makeatletter

\usepackage{color}

\makeatletter
\newcommand{\bee}{\begin{equation}}
\newcommand{\ee}{\end{equation}}
\newcommand{\beea}{\begin{eqnarray}}
\newcommand{\eea}{\end{eqnarray}}

\makeatother

\makeatother

\makeatother

\usepackage{babel}
\begin{document}
\begin{center}
\textbf{\Large{}Cosmological fluctuations: Comparing Quantum and Classical
Statistical and Stringy Effects}
\par\end{center}{\Large \par}

\begin{center}
\vspace{0.3cm}
 
\par\end{center}

\begin{center}
{\large{}S. P. de Alwis$^{\dagger}$ }
\par\end{center}{\large \par}

\begin{center}
Physics Department, University of Colorado, \\
 Boulder, CO 80309 USA 
\par\end{center}

\begin{center}
\vspace{0.3cm}
 
\par\end{center}

\begin{center}
\textbf{Abstract} 
\par\end{center}

The theory of cosmological fluctuations assumes that the pre-inflationary
state of the universe was the quantum vacuum of a scalar field(s)
coupled to gravity. The observed cosmic microwave background fluctuations
are then interpreted as quantum fluctuations. Here we consider alternate
interpretations of the classic calculations of scalar and tensor power
spectra by replacing the Bunch-Davies quantum vacuum with a classical
statistical distribution, which may have been the consequence of a
pre-inflationary process of decoherence as in the quantum cosmology
literature. Mathematically they are essentially identical calculations.
However if one takes the latter interpretation then one might replace
the Planck length by for instance the fundamental length scale of
string theory. In particular this changes the relation between the
scale of inflation and the scalar power spectrum but leaves the parameter(s)
characterizing the bi-spectrum unchanged at leading order. Differences
will occur however at higher order in the loop expansion. We also
discuss the relation to theories with low sound speed and/or a period
of dissipation during inflation (warm inflation).

\begin{center}
\vspace{0.3cm}
 
\par\end{center}

\vfill{}

$^{\dagger}$ dealwiss@colorado.edu

\eject

\section{Introduction}

The theory of cosmological fluctuations\footnote{See for instance \citep{Dodelson:2003ft}\citep{Mukhanov:2005sc}\citep{Weinberg:2008zzc}. }
is considered to be one of the crowning achievements of theoretical
cosmology. Given a model for inflationary cosmology this theory enables
one to calculate the measured scalar and tensor fluctuation spectra.
In particular the standard theory appears to tell us the absolute
value of the scale of inflation. In fact if the value of the ratio
of tensor to scalar power reported by BICEP2 \citep{Ade:2014xna}
had held up to scrutiny \citep{Ade:2015tva}, the theory would imply
that we are effectively looking at energy scales as large as $10^{16}GeV$,
a scale that is practically the same as the Grand Unification (GUT)
scale of particle physics. The latter is only two orders of magnitude
below the Planck scale - a scale at which quantum gravity effects
are necessarily of order one.

However this particular conclusion of the theory depends crucially
on the absolute normalization of the Fourier modes of the gravitational
tensor and scalar field modes. This is fixed by assuming that the
initial state of the universe just prior to inflation is the quantum
vacuum of free quantum fields - essentially an infinite product of
harmonic oscillator ground state wave functions, i.e. the so-called
Bunch-Davies vacuum. This is in effect the same as assuming that the
short wave length modes are picked from a particular Gaussian distribution
which involves $\hbar$ explicitly (see section \ref{sec:Comparison-with-cosmological})\footnote{For a discussion of the relation of quantum to classical distributions
that is in the same spirit as this paper see \citep{Brustein:2002zn}.}. Indeed this is the only occurrence of the quantum of action in the
entire theory of Gaussian cosmological fluctuations. The starting
point of our investigation is simply the observation that the standard
calculation of power spectra bi-spectra etc., are completely equivalent
mathematically to computing correlation functions in statistical mechanics
with a Gaussian measure and then evolving them in time using classical
Hamiltonian evolution rather than Heisenberg's equations. In the following
we will argue that this is as well motivated a assumption for the
physics of the initial state of inflation as the standard one, if
for example this distribution came about as the result of decoherence
of some pre-inflationary quantum cosmological state or an initial
quantum string state.

\subsection{Quantum field theory or classical statistical mechanics. }

What is truly peculiar to quantum behavior is a) the quantization
of energy $E=\hbar\omega$ and momentum $p=\hbar k$ for a wave of
angular frequency $\omega$ and wave number $k$, b) the quantization
of angular momentum - especially spin, c) long range correlations
of EPR type signaling the entanglement of quantum states. These are
properties that would be very hard if not impossible to reproduce
by appealing to classical statistical distributions. None of these
crucial properties of quantum behavior is however tested in the cosmological
observations. 

In view of this, one may ask whether there is a test at least of the
relevance of $\hbar$ to the calculation. If we start from the interpretation
of the initial configuration as a classical statistical distribution
the factor of $\hbar$ may be replaced by some arbitrary factor with
the same units. Consequently the relation between the power spectrum
and the scale of inflation will involve an undetermined factor. This
much of course should be rather obvious and is probably known to physicists
who have thought about these issues. Then the question is whether
there is some measurement in observation cosmology that can test (at
least) for the occurrence of Planck's constant in the data.

Moreover the possibility that the initial configuration is described
by a statistical distribution determined for instance by string theory,
has not been considered hitherto as far as we know. What is meant
here is not the procedure that has been followed in works such as
\citep{Martin:2000xs,Niemeyer:2001qe,Kaloper:2002uj,Easther:2002xe,Danielsson:2002kx},
where the initial state is still regarded as a pure quantum mechanical
state and the string theory effects are just viewed as corrections
to the background. Instead we argue that, since the standard calculations
of the inflaton fluctuations may actually be replaced by an initial
configuration which is Gaussian distributed with the width of distribution
governed by $\hbar$, one may consider replacing this by some other
constant with the same dimensions. For instance if the initial distribution
is the result of some pre-inflationary process of decoherence of some
initial (possibly stringy) pure state, which may have been the quantum
mechanical state of the multiverse, it is plausible that this initial
distribution is determined by stringy effects. String theory (unlike
quantum field theory) has a natural fundamental length scale $l_{s}$
which may be defined in terms of the Regge slope $\alpha'$ as $l_{s}^{2}=2\pi\alpha'$.
Together with the gravitational coupling constant $\kappa^{2}$ we
can then define (having set $c=1$) a fundamental unit of action $l_{s}^{2}/\kappa^{2}$.
Note that this is a ratio of two classical constants. Thus if string
theory is the fundamental theory of the universe, one might consider
as an alternative to the usual assumption for the initial configuration,
the possibility that it is a Gaussian distribution involving this
unit of action\footnote{If the compactification scale is the string scale and the string coupling
$g_{s}=1$, then these two units of action may be identified. However
if this is the case we are no longer in the weak coupling large volume
regime in which one is able to analyze the cosmology and phenomenology
of string theory.} rather than $\hbar$.

\subsection{Quantum cosmology issues and string theory}

In fact it seems unlikely that the standard argument for the initial
state of the inflationary cosmology being the Bunch Davies vacuum
is valid in the context of string theory. Within the context of a
scalar field theory coupled to gravity, the analysis of quantized
perturbations in the inflationary background is performed under the
assumption that the pre-inflationary primordial state of the universe
continues to be described by this theory - i.e. all the way back to
the initial singularity at the scale factor $a=0$. An argument for
the choice of this vacuum has been given in \citep{Maldacena:2002vr}
using contour rotation in an imaginary time direction. This essentially
corresponds to the choice of the Hartle-Hawking ``no boundary''
proposal for the wave function of the universe which vanishes at the
origin $a=0$. 

There are two issues that need to be considered in connection with
this reasoning. Firstly, it appears that this wave function (in contrast
to the so-called tunneling wave function), has very low probability
for leading to inflationary dynamics (see for instance \citep{Calcagni:2015vja}).
Secondly, and more importantly, if the fundamental theory that describes
the universe is string theory, then this simple picture is unlikely
to be the whole story particularly in the pre-inflationary stage.
It is far more likely to have been some primordial stringy state,
and the stage before inflation may have been one with a primordial
distribution of the decay products of string and Kaluza-Klein states.
Unfortunately in the absence of a solid theoretical construction of
such an initial state, all we can do is to parametrize our ignorance
with some simple ansatz. Here we choose to pick the initial state
from a Gaussian statistical distribution. We note that choosing the
width of the distribution to correspond to the free field theory one
gives us the usual story, while choosing it to correspond to one that
might plausibly arise from string theory, gives us different results
which may then be compared to the usual ones.

If the physics of the initial universe is governed by string theory,
then there is a further reason to think that the sort of conjecture
described in the previous paragraph may be reasonable. This is because
at low energies even at the classical level (i.e. zero string loop
level), one expects higher derivative ($R^{2}$ etc.) terms in the
effective action. These will lead to terms in the effective stress
tensor that will be larger than those that come from quantum effects
in quantum field theory. In fact as is well known (and discussed in
detail below) semi-classical string theory has a double expansion
- the $\alpha'$ expansion as well as the semi-classical string loop
expansion. It is the latter which corresponds to the the standard
calculations of QFT in curved backgrounds as discussed for instance
in the classic text book by Birrell and Davies \citep{Birrell:1982ix}.
The former is a purely stringy effect and is usually not considered
in cosmological discussions. One of the aims of this paper is to discuss
the consequences for the theory of cosmological fluctuations, of the
leading terms in the classical string theory $\alpha'$ expansion
assuming that this is related to the stringy modification of the initial
state that we described earlier. These as we will see below, are actually
larger than the terms which may be identified with the standard QFT
calculation.

To summarize: if one starts with the pre inflationary state as the
Bunch-Davies quantum vacuum one has to then explain how this led to
the statistical distribution that one actually does observe. Indeed
the process of decoherence may have actually happened before the inflationary
stage set in, from a primordial state in quantum cosmology (a solution
of the Wheeler deWitt equation) as discussed for instance in \citep{Barvinsky:1998cq}
and the references therein\footnote{Similar issues have also been discussed in a paper \citep{Maldacena:2015bha}
that appeared after the first version of the present work.}. The starting point for the usual calculations are then the classical
statistical distribution established by this process. But the process
itself and indeed the original quantum state which then decohered
to lead to the statistical distribution is irrelevant as far as observational
cosmology is concerned. The present paper is thus a modest attempt
to consider whether one can at least test the particular unit (i.e.
$\hbar$) of action that goes into this distribution, for instance
distinguishing it from a string theory derived unit.

\subsection{Units }

We work with units where the velocity of light $c=1$ (so that a unit
of time is equivalent to a unit of length $T=L$) but do not set $\hbar=1$
so that we keep independent units of length $L$ and mass $M$. Also
as usual we define $\kappa^{2}\equiv8\pi G$. Note that $[\hbar]=ML$
and $[\kappa^{2}]=M^{-1}L$ and the Planck length $l_{P}$ defined
by $l_{P}^{2}=\kappa^{2}\hbar$ has units of length while the Planck
mass $M_{P}$ defined by $M_{P}^{2}=\hbar/\kappa^{2}$ has dimensions
of mass. In quantum field theory (for instance the standard model
coupled to gravity treated semi-classically), the Planck length is
a derived quantity and is usually regarded as a length scale that
goes to zero in the classical limit $\hbar\rightarrow0$. In string
theory on the other hand there is an independent fundamental length
$l_{s}$ defined as the scale of the 2D sigma model\footnote{Unlike a standard action the string sigma model action has dimensions
of $L^{2}$ rather than $ML$ since the field is the coordinate in
the ambient space. So the functional integral is defined by introducing
a fundamental length scale $l_{s}$.}. In particular the loop expansion of the gravitational constant will
take the form
\begin{equation}
\frac{1}{2\kappa^{2}}=\frac{1}{2\kappa_{0}^{2}}+\frac{\hbar}{l_{s}^{2}}f(\frac{\kappa_{0}^{2}\hbar}{l_{s}^{2}})\label{eq:kappaexpansion}
\end{equation}
where $\kappa$ is the physical (i.e. renormalized) gravitational
constant $\kappa_{0}$ is the bare constant and $l_{s}$ may be naturally
identified with the string scale in string theory but is an arbitrary
short distance cutoff in QFT. In the rest of the paper the gravitational
constant is taken to be the physical constant $\kappa$.

Note also that we use the mostly positive metric convention and the
Ricci tensor is defined as $R_{jk}=R_{\,jik}^{i}$.

\section{Inflationary fluctuations}

\subsection{Review }

Inflationary cosmology\footnote{I've followed closely the discussion in the review \citep{Kinney:2009vz}
in this section. } is usually formulated in terms of a theory of a scalar field coupled
to gravity with the action,
\begin{eqnarray}
S & = & \frac{1}{2\kappa^{2}}\int d^{4}x\sqrt{g}R+S_{m},\label{eq:Sclassical}\\
S_{m} & = & -\int d^{4}x\sqrt{g}(\frac{1}{2}g^{\mu\nu}\partial_{\mu}\phi\partial_{\nu}\phi+V(\phi)).\label{eq:Sm}
\end{eqnarray}
Note that $[S]=ML,\,[g_{\mu\nu}]=L^{0}M^{0},\,[\phi]=M^{1/2}L^{-1/2}$.

The classical Einstein equation for this system is
\begin{eqnarray}
G_{\mu\nu} & = & \kappa^{2}T_{\mu\nu},\label{eq:Eclassical}\\
T_{\mu\nu} & \equiv & -\frac{2}{\sqrt{g}}\frac{\delta S_{m}}{\delta g^{\mu\nu}}=\partial_{\mu}\phi\partial_{\nu}\phi-\frac{1}{2}g_{\mu\nu}(\partial_{\lambda}\phi\partial^{\lambda}\phi+V(\phi)).\label{eq:Tclassical}
\end{eqnarray}

Since the system is generally covariant the Einstein equation implies
the conservation of the stress tensor and when there is only one scalar
field it also implies the equation of motion
\begin{equation}
\nabla^{2}\phi=\frac{1}{\sqrt{g}}\partial_{\mu}\sqrt{g}g^{\mu\nu}\partial_{\nu}\phi=-\frac{\partial V}{\partial\phi}.\label{eq:phieofm}
\end{equation}

For an FRW (homogeneous isotropic) background\footnote{We've assumed for simplicity that the three curvature is zero.}
$g_{\mu\nu}={\rm diagonal}(-1,a^{2}(t)\delta_{ij}),\,i,j=1,2,3;\,\phi=\phi(t)$,
we have two independent equations,
\begin{eqnarray}
H^{2} & \equiv & \left(\frac{\dot{a}}{a}\right)^{2}=\frac{1}{3}\kappa^{2}(\frac{1}{2}\dot{\phi}^{2}+V(\phi)),\label{eq:F1}\\
\dot{H} & = & -\frac{1}{2}\kappa^{2}\dot{\phi}^{2}.\label{eq:F2}
\end{eqnarray}
 Inflation requires a period of accelerated expansion which leads
to the so-called slow roll conditions 
\begin{equation}
\epsilon\equiv-\frac{\dot{H}}{H^{2}}\ll1,\,\eta\equiv-\frac{\ddot{H}}{2H\dot{H}}\ll1.\label{eq:epsiloneta}
\end{equation}
For the single inflaton case above this translates into conditions
on the potential: 
\begin{equation}
\epsilon=\frac{1}{2\kappa^{2}}\left(\frac{\partial_{\phi}V(\phi)}{V(\phi)}\right)^{2}\ll1,\eta=\frac{1}{\kappa^{2}}\left[\frac{\partial_{\phi\phi}V(\phi)}{V(\phi)}-\frac{1}{2}\left(\frac{\partial_{\phi}V(\phi)}{V(\phi)}\right)^{2}\right].\label{eq:epsiloneta2}
\end{equation}
To analyze fluctuations around this background one typically goes
to conformal coordinates in which the background metric takes the
form $ds^{2}=a^{2}(\tau)(-d\tau^{2}+\boldsymbol{dx}^{2})$ taking
spatial curvature to be zero. The standard procedure is to impose
canonical commutation relations on the scalar field after writing
$\phi(\tau,\boldsymbol{x})=\phi(\tau)+\delta\phi(\tau,\boldsymbol{x})$
where the first term is the classical background field. So one expands
\begin{equation}
\delta\phi(\tau,\boldsymbol{x})=\frac{1}{(2\pi)^{3/2}}\int d^{3}k(\phi_{k}(\tau)b_{\boldsymbol{k}}e^{i\boldsymbol{k.x}}+h.c.)\label{eq:deltaphi}
\end{equation}
Writing $\phi_{k}(\tau)=a^{-1}(\tau)u_{k}$ we have from \eqref{eq:phieofm}
(with $X'\equiv\frac{dX}{d\tau}$) in the slow roll approximation,
\begin{equation}
u''_{k}+(k^{2}-\frac{a''}{a})u_{k}=0.\label{eq:ukeqn}
\end{equation}
Now the standard argument is to identify the fluctuation spectrum
of $\phi$, with the quantum fluctuations of an essentially free field
in the vacuum state. Observing that for $k^{2}\gg\frac{a''}{a}$ one
has in effect plane waves, and the physics is that of Minkowski space,
so one can follow the standard procedure for flat space quantization.

The scalar field is taken to be canonically quantized i.e. $[\phi(\boldsymbol{x},\tau),\pi(\boldsymbol{y},\tau)]=i\hbar\delta^{3}(\boldsymbol{x}-\boldsymbol{y})$,
with the other two commutators vanishing. If the solutions to \eqref{eq:ukeqn}
are normalized with the usual Klein-Gordon norm (i.e. $u_{k}u^{*'}-u_{k}^{*}u'_{k}=i$)
then the operators $b_{\boldsymbol{k}}$ satisfy the relations 
\[
[b_{\boldsymbol{k}},b_{\boldsymbol{k'}}^{\dagger}]=\hbar\delta^{3}(\boldsymbol{k}-\boldsymbol{k'}),
\]
with the other commutators vanishing. It is convenient to define now
the scalar power spectrum although this is not what is physically
relevant (i.e. related to the temperature fluctuations). This is defined
as 
\begin{equation}
P(k,\tau)=\frac{\hbar k^{3}}{2\pi^{2}}|\phi_{k}(\tau)|^{2}\label{eq:Pphi}
\end{equation}
in terms of which the autocorrelation function of the scalar field
fluctuation (in a spatially translational invariant background) is
\begin{equation}
<\phi(\tau,\boldsymbol{x})\phi(\tau,\boldsymbol{y})>=\int\frac{dk}{k}P(k)\frac{\sin k|\boldsymbol{x}-\boldsymbol{y}|}{k|\boldsymbol{x}-\boldsymbol{y}|}\label{eq:phiphi}
\end{equation}
The power spectrum of the scalar curvature fluctuation which is related
to the measured temperature fluctuation) is then given as (with $N$
being the number of e-foldings regarded as a function of $\phi$ with
$dN=Hdt$) 
\begin{equation}
P_{{\cal R}}(k)=\left(\frac{\delta N}{\delta\phi}\right)^{2}P(k)=\frac{\kappa^{2}}{2}\frac{P(k)}{\epsilon}.\label{eq:PR}
\end{equation}
Note that this is independent of the normalization of $\phi$ as it
should be, since it is directly related to a measurable physical effect. 

For short wave lengths $k\tau\gg1$ as in the above discussion the
normalized solution (consistent with the Lorentz invariant plane wave
solution for $\phi$) to \eqref{eq:ukeqn} is $ $$u_{k}=e^{ik\tau}/\sqrt{2k}$
and 
\begin{equation}
P(k,\tau)=\frac{\hbar}{4\pi^{2}}\frac{k^{2}}{a^{2}(\tau)}=\frac{1}{4\pi^{2}}\frac{\hbar}{\lambda_{{\rm physical}}^{2}(\tau)}.\label{eq:Pkshort}
\end{equation}
For constant $\epsilon$ it is possible to find the exact solutions
to \eqref{eq:ukeqn} and the solution that asymptotes to the Minkowski
solution for short wave lengths is 
\begin{equation}
u_{k}(\tau)=\frac{1}{2}\sqrt{\frac{\pi}{k}}\sqrt{-k\tau}H_{\nu}^{(1)}(-k\tau),\label{eq:usoln}
\end{equation}
where $H_{\nu}^{(1)}$ is the Hankle function of the first kind and
\[
\nu=\frac{3-\epsilon}{2(1-\epsilon)}.
\]
In the long wave length limit we have 
\begin{equation}
[P(k,\tau)]^{1/2}=\hbar^{1/2}2^{\nu-3/2}\frac{\Gamma(\nu)}{\Gamma(3/2)}(1-\epsilon)\frac{H}{2\pi}\left(\frac{k}{a(\tau)H(1-\epsilon)}\right){}^{\frac{3}{2}-\nu}.\label{eq:Pkexact}
\end{equation}
For the exactly deSitter case ($\epsilon=0)$
\begin{equation}
P(k,\tau)=\hbar\left(\frac{H}{2\pi}\right)^{2}.\label{eq:PkdS}
\end{equation}

There is a similar formula for the tensor power spectrum. Defining
the tensor fluctuations around the FRW background as 
\begin{equation}
\delta g_{ij}\equiv a^{2}h_{ij}=a^{2}(h_{+}e_{ij}^{+}+h_{-}e_{ij}^{{\rm x}}),\,h_{\lambda}(\boldsymbol{k})=\frac{1}{(2\pi)^{3}}\int d^{3}xe^{i\boldsymbol{k}.\boldsymbol{x}}h_{\lambda}(\boldsymbol{x}).\label{eq:hdefn}
\end{equation}
Here $e_{ij}^{\lambda},\,\lambda=+,{\rm x}$ is the polarization 3-tensor
satisfying 
\begin{eqnarray}
e_{ij}^{+} & = & e_{ji}^{{\rm x}},\,k^{i}e_{ij}^{\lambda}=0,\,e_{ii}=0,\,e_{ij}^{\lambda}(-\boldsymbol{k})=e_{ij}^{\lambda}(\boldsymbol{k})^{*}\label{eq:poltensor}\\
\sum_{\lambda}e_{ij}^{\lambda*}e_{ij}^{\lambda} & = & 4.\label{eq:polsum}
\end{eqnarray}
From the Einstein action we then have in the Gaussian approximation
the action for the tensor fluctuations
\begin{equation}
\Delta S=\frac{1}{2\kappa^{2}}\int\sqrt{g_{(0)}}g_{(0)}^{\mu\nu}\frac{1}{2}\partial_{\mu}h_{ij}\partial_{\nu}h_{ij}\label{eq:haction}
\end{equation}
This is essentially a sum of four free scalar fields so that defining
$v_{k}\equiv\frac{1}{\sqrt{2}\kappa}a(\tau)h_{k}$ in analogy with\eqref{eq:deltaphi}
(and the line below it) we have for the power spectrum in tensors,
\begin{equation}
P_{T}(k)=\frac{k^{3}}{2\pi^{2}}\sum_{\lambda}|h_{\lambda k}|^{2}=8\kappa^{2}P(k)=8\hbar\kappa^{2}\left(\frac{H}{2\pi}\right)^{2}\left(\frac{k}{aH}\right)^{-2\epsilon}.\label{eq:PT}
\end{equation}
From \eqref{eq:Pkexact} and the above we have 
\begin{equation}
r\equiv\frac{P_{T}}{P_{{\cal R}}}|_{k=aH}=16\epsilon=-8n_{T}\label{eq:ratio}
\end{equation}
where we've parametrized $P_{T}\propto k^{n_{T}}$.

\subsection{Quantum effects in the stress tensor}

In this subsection it will convenient to use natural units $c=\hbar=1.$ 

The energy density gets a contribution from the inflaton quantum fluctuations:
\begin{equation}
<0|T_{00}^{\phi}|0>=a^{-2}(\tau)<0|T_{\tau\tau}^{\phi}|0>=\frac{1}{2}a^{-2}\int\frac{d^{3}k}{(2\pi)^{3}}(|\phi'_{k}|^{2}+k^{2}|\phi_{k}|^{2})\label{eq:Ttau}
\end{equation}
There is also a contribution from graviton fluctuations:
\[
<0|T_{00}^{h}|0>=8<0|T_{00}^{\phi}|0>.
\]
The quantum energy density may be written as a sum of three terms
in the different wave length regimes.
\begin{equation}
<0|T_{00}|0>=\int_{k_{I}\gg\tau^{-1}}^{k_{UV}}\frac{dk}{k}\frac{k^{4}}{4\pi^{2}a^{4}(\tau)}+\int_{k_{IR}=aH}^{k_{I}}\frac{dk}{k}\frac{k^{2}}{a^{2}(\tau)}P(k)+\int_{0}^{k_{IR}=aH}\frac{dk}{k}\frac{k^{2}}{a^{2}(\tau)}P(k).\label{eq:Too}
\end{equation}
 The last term above just gives a contribution (assuming exact deSitter
$\epsilon=0$ for simplicity) corresponding to the Gibbons-Hawking
temperature: 
\begin{equation}
<T_{00}>_{IR}\equiv\frac{4\pi^{2}}{2}\left(\frac{H}{2\pi}\right)^{4}=2\pi^{2}T_{dS}^{4}\label{eq:TIR}
\end{equation}
and is negligible compared to the classical energy during inflation
$<T_{00}>_{IR}\ll3H^{2}M_{P}^{2}\simeq V(\phi)$ since we must necessarily
have $H\ll M_{P}$ for the validity of the EFT. 

The contribution in the deep UV however is as usual divergent. To
evaluate it we impose a cutoff at some comoving scale $k_{UV}$ and
evaluate the short distance contribution from $k^{2}\gg a^{''}/a$
by using the Minkowski (BD vacuum) modes $\phi_{k}=a^{-1}u_{k}=a^{-1}e^{ik\tau}/\sqrt{2k}$.
\begin{equation}
<0|T_{00}^{\phi}|0>_{{\rm uv}}=\frac{1}{2a^{2}}\int^{k_{UV}}\frac{d^{3}k}{(2\pi)^{3}}\frac{2k^{2}}{2k}\sim\frac{k_{UV}^{4}}{16\pi^{2}a^{4}}=\frac{k_{UV}^{(phy)4}}{16\pi^{2}}\label{eq:Tuv}
\end{equation}
This must necessarily be smaller than the classical potential energy
density at the onset of inflation for the validity of inflationary
cosmology\footnote{The constraints coming from this for different inflationary scenarios
and how they can be mitigated in supersymmetric scenarios will be
discussed in a separate publication \citep{Burgess:2014cb}. } so that 
\begin{equation}
k_{UV}^{4}<192\pi^{2}H^{2}M_{P}^{2}\label{eq:nonsusybound}
\end{equation}

In a supersymmetric theory on the other hand this quartic divergence
will be absent and (for SUSY broken at a gravitino mass scale $m_{3/2}$
we have instead of \eqref{eq:nonsusybound} the relation
\begin{equation}
k_{UV}^{2}<192\pi^{2}\frac{H^{2}}{m_{3/2}^{2}}M_{P}^{2}.\label{eq:susybound}
\end{equation}

On the other hand at late times (i.e. after many e-folds of inflation),
the UV contribution (assuming a fixed physical cutoff at the onset
of inflation), will have been inflated away and only the last contribution
in \eqref{eq:Too} will survive. It is this that will be compared
with the string theory contribution below.

\section{String theory expansion}

In this section we revert back to units in which $c=1$ but $\hbar$
is not set equal to unity. 

In quantum field theory the perturbative expansion is an expansion
in the number of loops with $\hbar$ serving as a loop counting parameter.
The quantum effective (1PI) action has the expansion,
\begin{equation}
\Gamma(g_{\mu\nu},\phi,\hbar)=\Gamma_{_{0}}+\hbar\Gamma_{1}+\hbar^{2}\Gamma_{2}+\ldots=\sum_{l=0}^{\infty}\hbar^{l}\Gamma_{l},\,\Gamma_{0}=S,\label{eq:GammaFT}
\end{equation}
$S$ being the classical action.

In perturbative string theory \citep{Green:1987sp}\citep{Polchinski:1998rq}
each loop order is defined through the functional integral $Z=\int[dX]\exp\{-I\}$
over the embedding coordinates $X(\sigma)$ defining the world sheet
in the ambient space, weighted by the sigma model (dimensionless)
action
\[
I=\frac{1}{2\pi\alpha'}\int d^{2}\sigma\sqrt{\gamma}\gamma^{ab}\partial_{a}X^{\mu}\partial_{b}X^{\nu}g_{\mu\nu}(X)+\ldots..
\]
The $l$th loop order is defined by the action on a Riemann surface
of genus $l$ and $\alpha'$ is the squared string length so that
$[\alpha']=L^{2}$. The loop counting parameter here is the string
coupling $g_{s}$. Although there is no non-perturbative background
independent formulation of string theory valid at arbitrarily high
scales, one can still construct a low energy effective action. 

However in order to get low energy four-dimensional physics we need
to compactify string theory on an internal manifold. For the purpose
of discussing inflation in four dimensions we assume that the volume
of this manifold is fixed at a value ${\cal V}$ in string units (i.e.
the physical volume is ${\cal V}(2\pi\sqrt{\alpha'})^{6}$). Then
we have the following standard relations between the four dimensional
gravitational coupling $\kappa$, Planck's constant $\hbar$ and the
string theory parameters - namely:
\begin{equation}
\hbar\kappa^{2}\equiv l_{P}^{2}=\frac{2\pi\alpha'}{{\cal V}}g_{s}^{2}=\frac{l_{s}^{2}g_{s}^{2}}{{\cal V}}\label{eq:hbarg}
\end{equation}
 It is important to note that both $\kappa^{2}$ and $l_{s}^{2}\equiv2\pi\alpha'$
are classical parameters. So the semi-classical expansion in $\hbar$
is equivalent (in string theory) to the expansion in terms of the
squared string coupling $g_{s}^{2}$. Of course the validity of this
expansion requires that the dilaton has been stabilized such that
$g_{s}^{2}<1$.

Now the long distance quantum effective action coming from string
theory has a double expansion defined as follows. We have the (quantum)
semi-classical expansion as before. i.e. we again have \eqref{eq:GammaFT},
but now the expansion is in terms of $g_{s}^{2}$ so we have
\begin{equation}
\Gamma(g_{\mu\nu},\phi,g_{s})=\Gamma_{_{0}}+g_{s}^{2}\Gamma_{1}+g_{s}^{4}\Gamma_{2}+\ldots=\sum_{l=0}^{\infty}g_{s}^{2l}\Gamma_{l}.\label{eq:GammaString}
\end{equation}
Each term in this expansion is given at long distances (compared to
the string scale $l_{s}$) as an expansion in powers of $l_{s}^{2}$.
Thus we may write schematically (keeping only pure metric dependent
terms at higher orders),
\begin{equation}
\Gamma_{l}=\frac{1}{2\kappa^{2}}\int d^{4}x\sqrt{g}[\delta_{l0}(R+2\kappa^{2}{\cal L}_{m})+l_{s}^{2}(R)_{l}^{2}+l_{s}^{4}(R)_{l}^{3}+\ldots]\label{eq:Gammal}
\end{equation}
In the above ${\cal L}_{m}$ is the classical matter lagrangian and
the notation $(R)_{l}^{n}$ represents all possible contractions of
curvatures and covariant derivatives to yield terms with $2n$ derivatives
of the metric, with dimensionless loop order dependent coefficients
(some of which may be zero), that are determined once the string theory
data are given. 

The gravitational equation of motion coming from \eqref{eq:GammaString}
$\partial\Gamma/\partial g_{\mu\nu}=0$, then takes the form (after
moving all string/quantum corrections to the RHS of the equation),
\begin{equation}
R_{\mu\nu}-\frac{1}{2}g_{\mu\nu}R=\kappa^{2}T_{\mu\nu}.\label{eq:Eeqn}
\end{equation}
Here the RHS is the effective stress-energy tensor coming from the
full quantum theory and is given by the double infinite series,
\begin{eqnarray}
\kappa^{2}T_{\mu\nu} & = & \kappa^{2}<\hat{T}_{\mu\nu}>=\kappa^{2}T_{\mu\nu}^{(m)}+l_{s}^{2}(R^{2})_{\mu\nu}(c_{0}^{0}+c_{2}^{0}\frac{1}{{\cal V}^{2/3}}+c_{4}^{0}\frac{1}{{\cal V}}+\ldots)+O(l_{s}^{4})\,{\rm classical}\nonumber \\
 &  & +g_{s}^{2}[l_{s}^{2}(R^{2})_{\mu\nu}(c_{0}^{1}+c_{2}^{1}\frac{1}{{\cal V}^{2/3}}+c_{4}^{1}\frac{1}{{\cal V}}+\ldots)+O(g_{s}^{2}l_{s}^{4})]\,{\rm 1-loop}\label{eq:Tmunu}\\
 &  & +{\rm 2-loop}+\ldots.\nonumber 
\end{eqnarray}
The first line represents the classical contributions to the stress
tensor including all classical string corrections. Note however that
$c_{0}^{0}\ne0$ only for type I and Heterotic strings. For type II
strings $c_{0}^{0}=0$ and the leading classical term is the $c_{2}^{0}$
term\footnote{This comes from a ``$R^{4}$'' term in the 10D long distance effective
action.}. The second line is the leading quantum correction $O(\hbar)$ and
so on. The subscripts on the $c$'s is half the number of derivatives
in the six compact space directions in the original 10D action, from
which the corresponding 4D term is extracted. The powers of ${\cal V}$
come from the contractions in the internal directions which scale
like $g^{ij}={\cal V}^{-1/3}\hat{g}^{ij}$ where $\hat{g}$ is a fiducial
metric normalized such that the volume of the internal space is $(2\pi)^{6}\alpha'^{3}$.
Correspondingly we expect the internal curvature in the hatted metric
to be $O(l_{s}^{-2})$. The superscripts on the $c$'s give the loop
order. In principle given a string theory and its compactification
data these coefficients are determined.

\section{Comparison with cosmological calculation\label{sec:Comparison-with-cosmological}}

The standard expressions for the scalar and tensor curvature fluctuation
are 
\begin{equation}
P_{{\cal R}}=\frac{\hbar\kappa^{2}}{2\epsilon}\left(\frac{H}{2\pi}\right)^{2}=\frac{g_{s}^{2}l_{s}^{2}}{2\epsilon{\cal V}}\left(\frac{H}{2\pi}\right)^{2},\,P_{T}=8\hbar\kappa^{2}\left(\frac{H}{2\pi}\right)^{2}=8\frac{g_{s}^{2}l_{s}^{2}}{{\cal V}}\left(\frac{H}{2\pi}\right)^{2},\label{eq:PRPT}
\end{equation}
 where in the second equalities in each of the above we have reexpressed
$\hbar\kappa^{2}$ in terms of the string coupling and length scale
using \eqref{eq:hbarg}. The corresponding contribution to the stress
tensor is given by (see \eqref{eq:TIR}) 
\begin{equation}
\kappa^{2}<T_{00}>\sim2\pi^{2}\hbar\kappa^{2}\left(\frac{H}{2\pi}\right)^{4}=2\pi^{2}g_{s}^{2}\frac{l_{s}^{2}}{{\cal V}}\left(\frac{H}{2\pi}\right)^{4},\label{eq:Toneloop}
\end{equation}
and can only come from the first term in the second line of \eqref{eq:Tmunu}
and is therefore equivalent to a one-loop string effect. However as
discussed in the previous subsection, string theory may also have
somewhat larger contributions corresponding to some of the terms in
the first (classical) line of \eqref{eq:Tmunu} as well the leading
term in the second line. In fact the first term in the first line
would be larger than \eqref{eq:Toneloop} by a factor ${\cal V}/g_{s}^{2}$
. However it should be emphasized that this term is present (i.e.
$c_{0}^{0}\ne0$ in \eqref{eq:Tmunu}) only for type I or Heterotic
strings where the volume cannot be larger than about a factor of 20
so this is only a factor $\gtrsim10$. In type II strings on the other
hand the volume could be much larger ${\cal V}>10^{3}$. In this case
the leading classical term in \eqref{eq:Tmunu} is the $c_{2}^{0}$
term and gives a contribution which is a factor ${\cal V}^{1/3}/g_{s}^{2}\gg10$
larger than \eqref{eq:Toneloop}.

To understand where such a contribution might come from in the context
of the usual arguments, we should revisit the assumptions for inflationary
initial conditions.

Suppose that the initial value (at $\tau\rightarrow\tau_{0}\gg k^{-1}$
for all relevant comoving scales $k$) of the field $\phi$ is Gaussian
distributed with a probability density
\begin{equation}
p(\phi)d\phi={\rm lim}_{\tau\rightarrow\tau_{0}}\exp\left[-\frac{1}{2}\int d^{3}x\int d^{3}y\phi(\boldsymbol{x},\tau)K(\boldsymbol{x},\boldsymbol{y};\tau)\phi(\boldsymbol{y},\tau)\right]d\phi\equiv e^{-\frac{1}{2}\phi.K.\phi}d\phi.\label{eq:gaussprob}
\end{equation}
All initial correlation functions are then given in terms of the two
point function and are computed from the generating formula
\begin{equation}
<e^{J.\phi}>\equiv e^{W[J]}=e^{W(0)}e^{\frac{1}{2}J.K^{-1}.J}.\label{eq:WJ}
\end{equation}
Here $K^{-1}.K=\int d^{3}yK^{-1}(\boldsymbol{x},\boldsymbol{y})K(\boldsymbol{y},\boldsymbol{z})=\delta^{3}((\boldsymbol{x}-\boldsymbol{z})$
and initial two point function is given by 
\begin{equation}
<\phi(\boldsymbol{x},-\infty)\phi(\boldsymbol{y},-\infty)>=K^{-1}(\boldsymbol{x},\boldsymbol{y};-\infty).\label{eq:twopoint}
\end{equation}
Now the crucial assumption of the theory of cosmological fluctuations
is that the initial probability distribution is the same as that corresponding
to a initial quantum mechanical state given by the free field (harmonic
oscillator) vacuum. This corresponds to choosing (after setting $a(\tau=\tau_{0})=1$
for convenience) 
\begin{equation}
K=\frac{2{\cal E}}{\hbar}=\frac{1}{\hbar}\int\frac{d^{3}k}{(2\pi)^{3}}e^{i\boldsymbol{k.(x-y})}2k\label{eq:Kstandard}
\end{equation}
Then we have from \eqref{eq:twopoint} for the initial value of the
two point function $ $
\begin{equation}
<\phi(\boldsymbol{x},\tau_{0})\phi(\boldsymbol{y},\tau_{0})>=\hbar\int\frac{d^{3}k}{(2\pi)^{3}}\frac{e^{i\boldsymbol{k.(x-y})}}{2k}=\hbar\int\frac{dk}{k}\frac{k^{3}}{2\pi^{2}}\frac{1}{2k}\frac{\sin k|\boldsymbol{x}-\boldsymbol{y}|}{k|\boldsymbol{x}-\boldsymbol{y}|}\label{eq:twopoinitial}
\end{equation}
This in fact is the initial value of the standard calculation (see
\eqref{eq:phiphi} and \eqref{eq:Pphi}) which gives 
\[
<\phi(\boldsymbol{x},\tau)\phi(\boldsymbol{y},\tau)>=\hbar\int\frac{dk}{k}\frac{k^{3}}{2\pi^{2}}|\phi_{k}(\tau)|^{2}\frac{\sin k|\boldsymbol{x}-\boldsymbol{y}|}{k|\boldsymbol{x}-\boldsymbol{y}|}
\]
 when the limit $\tau\rightarrow\tau_{0}$ is taken since $\phi_{k}(\tau)\rightarrow e^{-ik\tau}/\sqrt{2k}$
(recall that we set $a(\tau=\tau_{0})=1$). The point is that the
dependence on $\hbar$ and hence the supposed quantum nature of the
cosmological perturbations, just comes from the normalization derived
from the assumption that the initial distribution of short wave length
fluctuations is given by the product of quantum harmonic oscillator
ground state wave functions.

It would be nice to have some criterion for actually testing this
hypothesis. But in any case we should entertain also the possibility
that the initial state of inflation is simply a classical statistical
distribution given by \eqref{eq:gaussprob}. In fact using \eqref{eq:deltaphi}
the probability distribution \eqref{eq:gaussprob} becomes (after
using \eqref{eq:Kstandard} and averaging over the initial time $\tau_{0}$
(with $|k\tau_{0}|\gg1$) so as to get rid of the oscillatory pieces,
\begin{equation}
p(\phi)d\phi=\exp\left(-\frac{1}{\hbar}\int d^{3}kb_{\boldsymbol{k}}b_{\boldsymbol{k}}^{*}\right)\prod_{\boldsymbol{q}}db_{\boldsymbol{q}}\prod_{\boldsymbol{p}}db_{\boldsymbol{p}}^{*}.\label{eq:probhbar}
\end{equation}
The usual free quantum field theory calculation is thus completely
equivalent to the above classical distribution which gives \footnote{Note that $[b_{\boldsymbol{k}}]=M^{1/2}L^{2}$.}
\begin{equation}
<b_{\boldsymbol{k}}b_{\boldsymbol{q}}^{*}>=\hbar\delta^{3}(\boldsymbol{k}-\boldsymbol{q}),\,<b_{\boldsymbol{k}}b_{\boldsymbol{q}}>=<b_{\boldsymbol{k}}^{*}b_{\boldsymbol{q}}^{*}>=0.\label{eq:bcorrhbar}
\end{equation}
Here we will consider the consequences of assuming that the initial
distribution is defined by replacing the quantum unit of action $\hbar$
in \eqref{eq:Kstandard} by some other unit of action ${\cal A}$.
Now in classical physics there is no fundamental unit of action but
in string theory one can define such a unit, 
\begin{equation}
{\cal A}=\frac{l_{s}^{2}}{\kappa^{2}},\label{eq:stringactionunit}
\end{equation}
where $l_{s}$ is the string scale defined after \eqref{eq:hbarg}\footnote{Recall that the standard normalization corresponds to setting ${\cal A}=\hbar=\frac{l_{s}^{2}}{\kappa^{2}}\frac{g_{s}^{2}}{{\cal V}}.$}.
In this case the initial probability distribution is given again by
\eqref{eq:gaussprob}, but now with the kernel being given by
\begin{equation}
K=\frac{2{\cal E}}{{\cal A}}=\frac{\kappa^{2}}{l_{s}^{2}}\int\frac{d^{3}k}{(2\pi)^{3}}e^{i\boldsymbol{k.(x-y})}2k\label{eq:Kstring}
\end{equation}
corresponding to having the correlator $<b_{\boldsymbol{k}}b_{\boldsymbol{q}}^{*}>=\frac{l_{s}^{2}}{\kappa^{2}}\delta^{3}(\boldsymbol{k}-\boldsymbol{q})$.
Equation \eqref{eq:phiphi} (with \eqref{eq:Pphi}) is replaced by
\begin{equation}
<\phi(\boldsymbol{x},\tau)\phi(\boldsymbol{y},\tau)>=\frac{l_{s}^{2}}{\kappa^{2}}\int\frac{dk}{k}\frac{k^{3}}{2\pi^{2}}|\phi_{k}(\tau)|^{2}\frac{\sin k|\boldsymbol{x}-\boldsymbol{y}|}{k|\boldsymbol{x}-\boldsymbol{y}|}\label{eq:phiphinew}
\end{equation}
and the scalar field power spectrum (for simplicity we take $\epsilon=0$)
becomes,
\[
P(k)=\frac{l_{s}^{2}}{\kappa^{2}}\left(\frac{H}{2\pi}\right)^{2}.
\]
 The physical power spectra for scalar curvature and tensor fluctuations
is now,
\begin{equation}
P_{{\cal R}}=\frac{l_{s}^{2}}{2\epsilon}\left(\frac{H}{2\pi}\right)^{2},\,P_{T}=8l_{s}^{2}\left(\frac{H}{2\pi}\right)^{2},\,r\equiv\frac{P_{T}}{P_{{\cal R}}}=16\epsilon.\label{eq:PRPT-het}
\end{equation}
These power spectra are a factor ${\cal V}/g_{s}^{2}$ larger than
the standard values quoted in \eqref{eq:PRPT}. Correspondingly a
given observed power spectrum will imply a lower scale of inflation
(by a factor $g_{s}^{2}/{\cal V}$) compared to the standard result.
Also this power spectrum corresponds to a contribution to the stress
tensor at late times that is of the same order as the leading string
correction in line one of \eqref{eq:Tmunu}. 

Thus the initial conditions with $\hbar$ replaced by the unit of
action \eqref{eq:stringactionunit} seems to correspond to the situation
that one might obtain in Heterotic and type I string theories where
the coefficient $c_{0}^{0}\ne0$. On the other hand if the string
theory is type II then this coefficient is zero and the leading term
is $O(l{}_{s}^{2}/{\cal V}^{\frac{2}{3}})$. This would correspond
to an initial distribution with a kernel whose normalization factor
is given by ${\cal A}=\frac{l_{s}^{2}}{{\cal V}^{2/3}\kappa^{2}}$
rather than \eqref{eq:stringactionunit}. In this case the power spectra
become,
\begin{equation}
P_{{\cal R}}=\frac{l_{s}^{2}}{2\epsilon{\cal V}^{2/3}}\left(\frac{H}{2\pi}\right)^{2},\,P_{T}=8\frac{l_{s}^{2}}{{\cal V}^{2/3}}\left(\frac{H}{2\pi}\right)^{2}.\label{eq:PRPT-IIB}
\end{equation}

Going back to natural units $\hbar=1$ for simplicity, the bound on
the cutoff \eqref{eq:susybound} becomes 
\begin{equation}
k_{UV}^{2}<192\pi^{2}\frac{l_{P}^{2}}{l_{s}^{2}}\frac{H^{2}}{m_{3/2}^{2}}M_{P}^{2},\label{eq:kuvhet}
\end{equation}
 for the Heterotic case (i.e. with ${\cal A}=l_{s}^{2}/l_{P}^{2}$),
and 
\begin{equation}
k_{UV}^{2}<192\pi^{2}{\cal V}^{2/3}\frac{l_{P}^{2}}{l_{s}^{2}}\frac{H^{2}}{m_{3/2}^{2}}M_{P}^{2},\label{eq:kuvIIB}
\end{equation}
for the IIB case with ${\cal A}=l_{s}^{2}/{\cal V}^{2/3}l_{P}^{2}$.
In a string theory set up one expects the $k_{UV}=M_{KK}=M_{P}/{\cal V}^{2/3}$
giving the mild constraints
\begin{eqnarray*}
{\cal V} & > & \frac{1}{(192\pi^{2}g_{s}^{2})^{3}},\,\,\,{\rm Heterotic},\\
{\cal V} & > & \frac{1}{(192\pi^{2}g_{s}^{2})^{1/3}},\,\,{\rm IIB}.
\end{eqnarray*}
The modified scalar curvature and tensor power spectra imply corresponding
contributions to the stress tensor from higher derivative terms. In
particular they would imply the second term on the first line of \eqref{eq:Tmunu}.
As discussed before these classical string contributions to the stress
tensor would be larger than the quantum effects of the standard contribution,
and from a string theory stand point perhaps justify the alternative
initial state suggested in \eqref{eq:Kstring}.

We also point out that the quantum corrections to the stress tensor
at late times implied by the usual assumptions for the initial state,
are consistent with the string theory arguments discussed above, only
if the classical stringy corrections (the $c_{i}^{0}$ terms) in the
expansion for the effective stress tensor \eqref{eq:Tmunu} are all
absent. This is generically not the case in string theory though it
is not inconceivable that there may be compactifications that have
this property. Now one may ask why one should choose the precise formula
\eqref{eq:stringactionunit} as replacement for $\hbar$ in \eqref{eq:bcorrhbar}
and hence in \eqref{eq:phiphinew}. Obviously any numerical multiple
(say ${\cal A}\rightarrow\sigma{\cal A})$ will have the same dimensions
and the relations \eqref{eq:PRPT-het} will get multiplied by $\sigma$.
This just reflects the fact that at the classical level there is no
reason to prefer one value of the unit of action over another. However
from a string theory point of view this ambiguity is fixed for a given
compactified string theory. As we saw at the end of the last section
even the order of magnitude of the normalization factor will change
depending on the type of string theory that is being considered. In
the classical $\alpha'$ expansion terms in the first line of \eqref{eq:Tmunu},
once a particular string theory and its compactification data are
given, the coefficients of the curvature squared terms are determined.
This in turn fixes the ambiguity in the value of $\sigma$.

\section{Effective Field Theory of Inflation}

As we've discussed above if the initial configuration is governed
by a Gaussian fluctuation spectrum that is fixed in terms of say the
string scale rather than the Planck scale there would be a significant
difference in what the scale of inflation is for a given tensor to
scalar ratio $r$. Thus from \eqref{eq:PRPT-het} we have (we use
natural units in this section), 
\begin{eqnarray*}
H^{2} & = & 8\pi^{2}P_{{\cal R}}\epsilon M_{s}^{2}=8\pi^{2}P_{{\cal R}}\frac{r}{16}\frac{M_{P}^{2}}{{\cal V}},\,{\rm Heterotic/Type\,I}\\
H^{2} & = & 8\pi^{2}P_{{\cal R}}\epsilon M_{s}^{2}{\cal V}^{2/3}=8\pi^{2}P_{{\cal R}}\frac{r}{16}\frac{M_{P}^{2}}{{\cal V}^{1/3}},\,{\rm Type\,II}
\end{eqnarray*}
which implies that the scale of the inflationary potential is (using
the approximate Friedman equation $H^{2}\simeq V/3M_{P}^{2}$ and
the observed value $P_{{\cal R}}\simeq10^{-9}$), 
\begin{eqnarray}
V^{1/4} & = & (15r)^{1/4}\sqrt{\pi}10^{-2}\frac{M_{P}}{{\cal V}^{1/4}}\,{\rm Heterotic/Type\,I}\label{eq:potscale1}\\
V^{1/4} & = & (15r)^{1/4}\sqrt{\pi}10^{-2}\frac{M_{P}}{{\cal V}^{1/12}},\,{\rm Type\,II}\label{eq:potscale2}
\end{eqnarray}
The standard result is 
\begin{equation}
V^{1/4}==(15r)^{1/4}\sqrt{\pi}10^{-2}M_{P}.\label{eq:potscale}
\end{equation}
So if we replace the variance of the statistical distribution corresponding
to the standard result in the manner discussed above, the scale of
the potential can be significantly lower for large compactification
volumes - at least in the heterotic case.

On the other hand in bottom up approaches to inflationary fluctuations
it appears possible to have a small speed of sound that will also
result in a significantly different relation between the scalar spectrum
and the scale of inflation. We would like in the following to compare
and contrast the two cases below. We will also discuss situations
in which there are significant dissipative effects during inflation
i. e. the ``warm inflation'' scenario.

\subsection{Bottom up construction and observational consequences}

An effective field theory for inflationary perturbations has been
developed in \citep{Cheung:2007st,Cheung:2007sv}. This was constructed
by using the symmetries of the theory around an approximately de Sitter
background. The theory is first constructed in a unitary gauge where
the time coordinate is chosen so that the fluctuations of the inflation
$\delta\phi(x,t)=0$. Then the so-called Stueckelberg trick is used
to restore time diffeomorphism invariance by introducing the Goldstone
boson $\pi$ through the replacement $t\rightarrow t+\pi({\bf x},t)$,
with the transformation $\pi({\bf x},t)\rightarrow\pi({\bf x},t)-\xi^{0}({\bf x},t)$
under temporal diffeomorphisms $t\rightarrow t+\xi^{0}({\bf x},t),\,{\bf x}\rightarrow{\bf x}$.
This leads to the following effective Lagrangian density (for details
see the above references):
\begin{eqnarray}
{\cal L} & =\frac{1}{2}M_{P}^{2}R & +\epsilon M_{P}^{2}H^{2}\left(\dot{\pi}^{2}-\frac{1}{a^{2}}(\partial_{i}\pi)^{2}\right)+2M_{2}^{4}\left(\dot{\pi}^{2}+\dot{\pi}^{3}-\dot{\pi}\frac{1}{a^{2}}(\partial_{i}\pi)^{2}\right)\nonumber \\
 &  & -\frac{4}{3}M_{3}^{4}(\dot{\pi}^{3}+\ldots)\nonumber \\
 &  & +M_{4}^{4}(16\dot{\pi}^{4}+\ldots)+\ldots.\label{eq:BEFT}
\end{eqnarray}
The ellipses within parenthesis represent terms with at least two
spatial derivatives while the ellipses at the end are terms starting
with $\dot{\pi}^{5}$. Note that this effective Lagrangian is expected
to be valid \footnote{It is convenient to use natural units in this section.}
in the energy range $E_{{\rm mix}}\ll E\ll\Lambda$ where typically
$E_{mix}\sim\epsilon H$ is the mixing scale below which mixing with
gravity cannot be ignored, and $\Lambda$ given by 
\begin{equation}
\Lambda^{4}=16\pi^{2}\epsilon M_{P}^{2}H^{2}\frac{c_{s}^{5}}{1-c_{s}^{2}},\label{eq:cutoff}
\end{equation}
is the UV cutoff. This is determined by finding the scale at which
the theory violates unitarity (or by calculating the scale at which
the theory becomes strongly coupled). $c_{s}$ is the speed of sound
- defined by
\begin{equation}
2M_{2}^{4}\equiv(c_{s}^{-2}-1)\epsilon M_{P}^{2}H^{2}.\label{eq:sound}
\end{equation}

The Power spectrum in such a theory is given by
\begin{equation}
P_{{\cal R}}=\frac{l_{P}^{2}}{2\epsilon c_{s}}\left(\frac{H}{2\pi}\right)^{2},\,P_{T}=8l_{P}^{2}\left(\frac{H}{2\pi}\right)^{2}\,r\equiv\frac{P_{T}}{P_{{\cal R}}}=16\epsilon c_{s}.\label{eq:c_spower}
\end{equation}
Let us note now that the magnitude of the scalar power spectrum $P_{{\cal R}}\sim10^{-9}$
and the requirement that the validity of the effective theory implies
$H\ll\Lambda$ gives the lower bound \citep{Cheung:2007st} 
\begin{equation}
c_{s}\gg P_{{\cal R}}^{1/4}\sim0.003.\label{eq:csbound}
\end{equation}
In a theory with a small $c_{s}$ close to the above lower bound,
the scalar power spectrum computed with the usual quantum vacuum gives
a relation between the scalar power spectrum and the scale of inflation
which is similar to the expression that one gets with a theory in
which $c_{s}\simeq1$ but the initial configuration is a statistical
distribution governed by $l_{s}$ rather than $l_{P}$ (see eqn \eqref{eq:PRPT-het}).
In fact if $c_{s}\simeq l_{P}^{2}/l_{s}^{2}$ the two expressions
would be identical and for a given scale of inflation and $\epsilon$
will have identical power spectra. However contrary to eqn \eqref{eq:PRPT-het},
in theories with low $c_{s}$ the tensor spectrum is unaltered. Thus
the usual relation between the scale of inflation and $r$ will be
unaltered from \eqref{eq:potscale}.

\subsubsection{Dissipative effects}

Some authors \citep{Berera:1995ie,Berera:2008ar}\citep{LopezNacir:2011kk}
have proposed a scenario (``warm inflation'') in which the inflaton
field during the slow roll phase is coupled to another sector (a heat
bath) to which it loses energy and the fluctuations are effectively
due to classical thermal effects - the system has lost its memory
of the BD quantum vacuum The dissipation is characterized by an energy
scale $\gamma\gg H$ and a heat bath at a temperature $T\sim H$.
The power spectrum \eqref{eq:c_spower} is then 
\[
P_{{\cal R}}=\left(\frac{\sqrt{\pi\gamma H}T}{c_{s}^{2}H^{2}}\right)\frac{l_{P}^{2}}{2\epsilon c_{s}}\left(\frac{H}{2\pi}\right)^{2}\equiv\alpha_{{\rm warm}}\frac{l_{P}^{2}}{2\epsilon c_{s}}\left(\frac{H}{2\pi}\right)^{2},
\]
The first factor in parenthesis is that by which the power spectrum
is enhanced over and above the result\eqref{eq:c_spower} in the absence
of dissipation. The tensor spectrum is unchanged. 

Suppose now that we focus on a class of inflationary models such as
the Starobinsky model and closely related (``Starobinsky like'')
ones in which $r$ is (perhaps) unobservable small. In this case with
$\epsilon\sim1/N^{2}$ where $N$ is the number of e-folds of inflationary
expansion. Typically in the more realistic sy tring theoretic models
such as fibre inflation \citep{Cicoli:2008gp}, this is around 60
and $\epsilon\sim10^{-3}$. Small values of $c_{s}$ not that far
from the lower bound above would then give a scale for the potential
that could not be distinguished from \eqref{eq:PRPT-het} with ${\cal V}\sim c_{s}^{-1}$.
On the other hand even if $c_{s}\lesssim1$ but thermal effects are
large (i.e. $\gamma\gg H$), then we will again get a situation that
is similar to the one without such effects but with a modified initial
configuration. In other words a low sound speed and or large thermal
effects could not be distinguished from a initial statistical distribution
(with $c_{s}\lesssim1$ and no dissipation during inflation) if only
the scalar power spectrum is known.

\subsection{Top down issues - identification of the cutoff.}

While it is possible to find the scale at which a given EFT breaks
down (as in the above discussion) here we would like to identify the
physics above the cut off scale $\Lambda$. In principle this should
enable us to identify the arbitrary parameters $M_{i}^{4}$ in the
above EFT. Here we will assume that that the UV physics is described
by string theory. 

However in order to perform this matching one needs precise expressions
coming from string theory for the higher order operators in the EFT.
This may be possible in principle but in practice it is still a daunting
task and of course is also subject to the choice of compactification
data etc. Nevertheless the assumption that the UV theory is (compactified)
string theory will as we shall give some useful constraints on the
constants of the EFT. In effect what we will do is to discuss a toy
model with higher dimension operators that may plausibly come from
string theory. In other words we would like to compare the EFT of
the previous subsection with an EFT that (up to the above caveats)
may be identified with a string theoretic low energy action.

We consider a model of the form\footnote{The DBI inflation model (for a review with a discussion of possible
problems in realizing such a model from string theory see \citep{Chen:2010xka})
would be a special case of this.}
\begin{equation}
S=\frac{M_{P}^{2}}{2}\int d^{4}x\sqrt{g}(R+\ldots)+\int d^{4}x\sqrt{g}(P(X,\phi)+\ldots,\label{eq:RPaction}
\end{equation}
with $X=-\frac{1}{2}g^{\mu\nu}\partial_{\mu}\phi\partial_{\nu}\phi$
and $P$ is taken to be an arbitrary smooth function of $X$ and the
inflaton $\phi$ such that to leading order in an expansion in $X$,
$P(X)\sim X+V(\phi)+O(X^{2})$. The first set of ellipses represent
higher curvature terms and the second set involves higher (than first)
derivatives of the fields, which in general will be present in a string
theoretic model. In addition of course there will be other light fields
(after integrating out string modes and KK modes), but we will assume
that there is some string theoretic set up which results in single
field inflation, and that all these complications are not relevant
as far as comparison with the EFT of inflation is concerned. 

In any case our focus is just on identifying the physical cut off
that a string theoretic model would require. Ignoring these additional
terms one gets the class of models studied in \citep{Garriga:1999vw}\footnote{Some aspects of the comparison of this model with the EFT are discussed
in Appendix A of \citep{Cheung:2007sv}. However they do not discuss
the general expectation for the cosmological EFT coefficients from
a top down point of view.}. Our only point here is that the relevant scale of this theory should
be identified with the KK (or string) scale of string theory if this
sort of model is to make any sense at all. Comparing with the EFT
studied in the previous section will lead us to identify the cut off
and hence the sound speed and the other arbitrary parameters of that
discussion.

The stress tensor of this theory is given by the perfect fluid form
with the energy density $E=2XP,_{X}-P$ and pressure identified as
$P$ and the speed of ``sound'' $c_{s}$ in the system is given
by 
\[
c_{s}^{2}=\frac{dP}{dE}=\frac{P,_{X}}{P,_{X}+2XP,_{XX}}.
\]
The relevant equations of motion are 
\begin{eqnarray}
3M_{P}^{2}H^{2} & = & E\label{eq:kF1}\\
\dot{E}+3H(E+P) & = & 0\label{eq:kF2}
\end{eqnarray}
The last can be replaced by the equivalent form
\begin{equation}
2XP,_{X}=E+P=-2M_{P}^{2}\dot{H}=2\epsilon M_{P}^{2}H^{2}.\label{eq:kF3}
\end{equation}
In the flat FRW metric we have the solution $X=X_{0}=\frac{1}{2}\dot{\phi}_{0}^{2}$
with in particular $X_{0}P,_{X}|_{0}=\epsilon M_{P}^{2}H^{2}$. Suppose
the scale of the theory is $\Lambda$. This is to be identified with
the mass of the lowest mass state (lowest KK/string state) that has
been integrated out. Thus we will write 
\begin{equation}
x=\frac{X}{\Lambda^{4}},\,\hat{\phi}=\frac{\phi}{\Lambda},\,P=\Lambda^{4}\bar{P}(x,\hat{\phi})\label{eq:scaled-X}
\end{equation}

Introducing a fluctuation $\delta x$ around a background homogeneous
solution of the equations of motion $\hat{\phi}_{0}=\phi_{0}/\Lambda,\,x_{0}=X_{0}/\Lambda,$
we 
\begin{equation}
x=x_{0}(t)+2x_{0}(t)(\dot{\pi}+X_{\pi}),\,\pi\equiv\delta\phi/\dot{\phi}_{0},\,X_{\pi}\equiv\frac{\dot{\pi}^{2}}{2}-a^{-2}\frac{1}{2}(\partial_{i}\pi)^{2}\label{eq:pifluctuation}
\end{equation}
may write the Taylor expansion around $x_{0}$ in a power series in
$\delta x$, after rearranging in terms powers of $\partial\pi$ (and
rewriting $\bar{P}\rightarrow P$ for convenience),
\begin{eqnarray*}
P(x) & = & P(x_{0})+2x_{0}P'(x_{0})\dot{\pi}+2x_{0}P'(x_{0})\left(X_{\pi}+\frac{2x_{0}P''(x_{0})}{P'(x_{0})}\frac{\dot{\pi}^{2}}{2}\right)\\
 &  & +\frac{(2x_{0}P'(x_{0}))^{2}}{2}\frac{P''(x_{0})}{(P'(x_{0}))^{2}}2\dot{\pi}X_{\pi}+\frac{1}{3!}(2x_{0}P'(x_{0}))^{3}\frac{P'''(x_{0})}{P'(x_{0})^{3}}(\dot{\pi}^{3}+\ldots)\\
 &  & +\frac{1}{4!}(2x_{0}P'(x_{0}))^{4}\frac{P^{(4)}(x_{0})}{(P'(x_{0}))^{4}}(\dot{\pi}^{4}+\ldots)+\ldots.
\end{eqnarray*}
The ellipses represent in addition to quintic and higher powers in
$\partial\pi$ also terms with at least one non-derivative factor
$\pi$ as well as terms with more than one derivative acting on $\pi$,
which for the purpose of comparison with \eqref{eq:BEFT} we ignore.
Using \eqref{eq:kF3} to put
\begin{equation}
x_{0}P'(x_{0})=\frac{\epsilon M_{P}^{2}H^{2}}{\Lambda^{4}}\equiv\lambda,\,2\lambda\frac{P''(x_{0})}{P'^{2}(x_{0})}=c_{s}^{-2}-1,\label{eq:lambdacs}
\end{equation}
gives us the expansion
\begin{eqnarray}
P(x) & = & P(x_{0})+2\lambda\dot{\pi}+2\lambda\left(c_{s}^{-2}\frac{\dot{\pi}^{2}}{2}-\frac{1}{2}a^{-2}(\partial_{i}\pi)^{2}\right)\nonumber \\
 &  & +2\lambda(c_{s}^{-2}-1)\dot{\pi}X_{\pi}+\frac{4}{3}\lambda^{3}\frac{P'''(x_{0})}{(P'(x_{0}))^{3}}(\dot{\pi}^{3}+3\dot{\pi}^{2}X_{\pi}+\ldots)\nonumber \\
 &  & +\frac{2}{3}\lambda^{4}\frac{P^{(4)}(x_{0})}{(P'(x_{0}))^{4}}(\dot{\pi}^{4}+\ldots)+\ldots.\label{eq:TEFT}
\end{eqnarray}
Comparing with \eqref{eq:BEFT} we identify
\begin{eqnarray}
M_{2}^{4} & = & (c_{s}^{-2}-1)\frac{1}{2}\epsilon M_{P}^{2}H^{2}=\lambda\Lambda^{4}\frac{P_{0}^{''}}{(P'_{0})^{2}}=\epsilon M_{P}^{2}H^{2}\frac{P_{0}^{''}}{(P'_{0})^{2}}\label{eq:M2}\\
M_{3}^{4} & = & -\lambda^{2}\Lambda^{4}\frac{P_{0}'''}{(P_{0}')^{3}}=\lambda\epsilon M_{P}^{2}H^{2}\frac{P_{0}'''}{(P_{0}')^{3}},\,M_{4}^{4}=\lambda^{2}\epsilon M_{P}^{2}H^{2}\frac{P_{0}^{(4)}}{(P'_{0})^{4}},\ldots\label{eq:M3M4}
\end{eqnarray}
Note that \eqref{eq:M2} implies that a very small speed of sound
($c_{s}\ll1$) requires an anomalously large ratio $\frac{P_{0}^{''}}{(P'_{0})^{2}}\gg1$.
For the above expansion to make sense the parameter $\lambda$ must
be small and the generic ratio$P_{0}^{(n)}/(P_{0}^{'})^{n}$ should
not be anomalously large or small. These numbers are expected to be
$O(1)$. Of course this does not preclude one or more coefficients
being anomalously large of small. However from a microscopic point
of view such an anomalous coefficient needs to be justified!

Now we can identify the cutoff $\Lambda$ - the scale at which that
EFT violates unitarity \eqref{eq:cutoff} with the top down scale
of our string theoretic model. This gives $\lambda^{-1}=16\pi^{2}\frac{c_{s}^{5}}{1-c_{s}^{2}}$.
The requirement $\lambda<1$ then leads to the lower limit on the
sound speed, $c_{s}>0.3$6. 

As we saw at the end of the previous subsection, purely from the low
energy EFT point of view (see eqn \eqref{eq:cutoff}) one can have
extremely small values of $c_{s}$ provided the validity of the theory
is lowered correspondingly, and conversely if one takes $c_{s}$ arbitrarily
close to one, the cutoff can be taken arbitrarily large. However as
we have seen, from the top down point of view, the cutoff cannot be
made smaller than $\epsilon M_{P}^{2}H^{2}$ without violating the
expansion in powers of $\partial\pi$ on which the EFT is based. From
the top down point of view the arbitrary coefficients $M_{n}^{4}$
in the EFT of the previous subsection must generically become systematically
smaller - even if some of them may be anomalously large.

Furthermore in a complete 4D EFT description (including the reheating
phase) the energy density in inflation (i.e. the height of the potential)
must be restricted by the UV scale - i.e. the lowest scale of the
UV theory that has been integrated out. This is because most of this
energy density will get converted to kinetic energy at the end of
inflation, and the latter is necessarily bounded by for example the
KK scale in string theory. So we need to have also $M_{P}^{2}H^{2}\lesssim\Lambda^{4}$.
This then gives from \eqref{eq:cutoff} a stronger bound on $c_{s}$
namely
\[
c_{s}^{5}\gg\frac{1-c_{s}^{2}}{16\pi^{2}\epsilon}\simeq\frac{1}{16\pi^{2}\epsilon},
\]
where the last relation is valid for small $c_{s}$. 

Of course there may be special cases where due to some symmetry there
is a relation between the coefficients $P_{0}^{(r)}$. The only known
example of this is DBI inflation . This corresponds (in our notation\eqref{eq:scaled-X}
with $\bar{P}\rightarrow P$) to having $P=f^{-1}(\hat{\phi})\sqrt{1-f(\hat{\phi})x)},$
so that 
\begin{eqnarray}
x_{0}P_{0}^{'} & = & \frac{1}{2}x_{0}(1-fx_{0})^{-1/2}=\lambda\label{eq:DBI1}\\
M_{2}^{4} & = & -\epsilon M_{P}^{2}H^{2}\frac{1}{2}(c_{s}^{-2}-1)=-\epsilon M_{P}^{2}H^{2}\frac{f}{\sqrt{1-fx_{0}}}\label{eq:DBI2}\\
M_{n}^{4} & = & (-1)^{n-1}\lambda^{n-2}\epsilon M_{P}^{2}H^{2}\left(\frac{c_{s}^{-2}-1}{2}\right)^{n-1}\label{eq:DBI3}
\end{eqnarray}
Again for the expansions \eqref{eq:BEFT}\eqref{eq:TEFT} to be valid
clearly $c_{s}^{-2}-1\ll\lambda^{-1}$ and we can have small sound
speed $1\gg c_{s}\gg\sqrt{\lambda}$. Of course in this case one is
supposed to have an exact expression for the infinite sum. The former
can be valid for arbitrarily small $c_{s}$ the only restriction being
$fx_{0}=\lambda(c_{s}^{-2}-1)/(1+\lambda(c_{s}^{-2}-1))<1$ which
translates to $c_{s}<1$. Clearly this is a very special situation.

\subsubsection{General expectations for EFT coefficients}

Abstracting from the above discussion, we expect in an effective field
theory of single field inflation where the cutoff is set by some UV
complete theory such as string theory, the following form for the
Lagrangian.
\begin{eqnarray}
{\cal L} &  & =\frac{1}{2}M_{P}^{2}R+\epsilon M_{P}^{2}H^{2}\times\nonumber \\
 &  & \left[\left(\dot{\pi}^{2}-\frac{1}{a^{2}}(\partial_{i}\pi)^{2}\right)-(c_{s}^{-2}-1)\left(\dot{\pi}^{2}+\dot{\pi}^{3}-\dot{\pi}\frac{1}{a^{2}}(\partial_{i}\pi)^{2}\right)+\sum_{n=3}^{\infty}a_{n}\lambda^{n-2}(\dot{\pi}^{n}+n\dot{\pi}^{n-1}X_{\pi}+\ldots)\right]\nonumber \\
 &  & +\epsilon M_{P}^{2}H^{2}(\sum_{n=3}b_{n}\lambda^{n-2}H\pi(\partial\pi)^{n-1}+\ldots).\label{eq:GEFT}
\end{eqnarray}
The last line represents terms with at least one non-derivative factor.
Note that for a given order in the derivative expansion the terms
with non-derivative factors are suppressed by at least one power of
the slow roll parameter - i.e. $b_{n}\sim O(\epsilon)$. The main
point of the above however is that unlike what emerges from a purely
bottom up discussion \footnote{See for example the discussion of the four point function in \citep{Senatore:2010jy}. }
the top-down approach tells us that the higher point functions in
the interaction are generically suppressed systematically by powers
of $\lambda$. Thus unless one or more of the coefficients $a_{n}$
is unnaturally large, one would not expect for instance the four point
function (governed by $M_{4}^{4}$) to be large if, as is the case
observationally, the three point function is small. In the absence
of some argument in the microscopic theory for $a_{4}$ to be large,
one should not assume it to be the case.

What this means is that the observational constraints on the three
point function will imply that the four and higher point functions
will be smaller by powers of $\lambda$ compared to the three point
function. In the next section we will compare the results for higher
point point functions when initial conditions are changed as was discussed
in previous sections in this paper, and compare it to what is expected
from the standard initial conditions applied to situations with very
small speed of sound and or significant dissipation effects during
inflation.

\section{Higher point functions and quantum vs classical evolution}

As discussed above (and shown in more detail in Appendix I) the only
thing ``quantum'' about the usual calculation is the use of the
unit of action $\hbar$ in the normalization of the two point function.
Now we will address the question of whether the replacement of Planck's
unit of action by some other unit ${\cal A}$ will give rise to observable
consequences - for instance in the Bispectrum, Trispectrum or higher
point functions (for a review see \citep{Chen:2010xka}). As discussed
in detail in the appendix the only possible occurrence of the unit
of action is in the expression for the Wick contraction (the two point
function). 

The expectation value of the observable $A(t)$, a product of field
operators at the time t, is given by evaluating 
\begin{equation}
<\hat{A}(t)>=\int\Pi db_{\boldsymbol{k}}db_{\boldsymbol{k}}^{*}e^{-\frac{1}{\hbar}\int b_{\boldsymbol{q}}b_{\boldsymbol{q}}^{*}d^{3}q}A[b,b^{*};t],\label{eq:bb*integral}
\end{equation}
where as discussed in the Appendix the factor $A[b,b^{*};t]$ which
is usually evaluated using quantum operator equations of motion, can
be equally well evaluated (in terms of its initial value) by using
the classical evolution equation \eqref{eq:AIHVordered} or\eqref{eq:AIHVcomm}.
This evolution does not give rise to any factors of $\hbar$. All
such factors come from the correlators
\begin{equation}
<b_{\boldsymbol{k}}b_{\boldsymbol{q}}^{*}>=\hbar\delta^{3}(\boldsymbol{k}-\boldsymbol{q}).\label{eq:bbstar}
\end{equation}
Let us now use the notation explained in the Appendix where we write
$z^{i}$ for a field (for $i=1,\ldots,n$) or its conjugate variable
(for $i=n+1,\ldots,2n$). The interaction Hamiltonian (in the interaction
picture) is then at least cubic in the interaction picture fields
$z_{I}^{i}$ which obey free field equations of motion and can be
expressed in terms of the classical solution to the free field equations
and the $b_{\boldsymbol{k}}$'s. Suppose that we wish to compute \eqref{eq:bb*integral}
for an n-point function at equal times, $A=A_{n}$. Consider the $N$th
term of the expression for $A_{n}[b,b^{*}]$ in equation \eqref{eq:AIHVcomm}
of Appendix I, namely
\begin{eqnarray*}
\int_{t_{0}}^{t}dt_{1}\int_{t_{0}}^{t_{1}}\ldots\int_{t_{0}}^{t_{N-1}}dt_{N} & [H_{I}^{V}(z_{I}(t_{N}),t_{N}),[H^{V}(z_{I}(t_{N-1}),t_{N-1})[\ldots[H_{I}^{V}(z_{I}(t_{1}),t_{1}),A_{n,I}(t)],\ldots]]\\
 & \sim A_{n+N,I}+\ldots,
\end{eqnarray*}
since each commutator removes one factor of $z_{I}$ from $A_{n}$
and one factor from $H_{I}$ and adds two factors of $z$. $H_{I}$
in general has higher than third order terms in $z$ and the ellipses
represent terms which have more factors of $z$. However as we discussed
in detail in the previous section these are suppressed by powers of
$\lambda\equiv\epsilon M_{P}^{2}H^{2}/\Lambda^{4}$ where $\Lambda$
is the UV cutoff which in string theory is the KK scale (if not the
string scale). So let us ignore them, although of course in evaluating
the higher point functions they will be needed since they could become
competitive with loop terms involving lower order interactions. Then
for $n=2r+1,$ the only non-vanishing terms have $N$ odd. Writing
$N=2M+1$, the factor of $\hbar$ coming from evaluating the integral
on the RHS of \eqref{eq:bb*integral} is $\hbar^{(n+N)/2}=\hbar^{r+M+1}$.
Similarly for $n=2s$ the non-vanishing terms are proportional to
$\hbar^{s+P}$ where $N=2P$. Thus we have
\begin{eqnarray}
<\hat{A}_{2r+1}(t) & \sim & \hbar^{r+1}(a_{r1}+a_{r2}\hbar+\ldots+a_{rM}\hbar^{M}+\ldots)\label{eq:Ahatodd}\\
<\hat{A}_{2s}(t)> & \sim & \hbar^{s}(b_{s0}+b_{s1}\hbar+\ldots+b_{sP}\hbar^{P}+\ldots)\label{eq:Ahateven}
\end{eqnarray}
Note that for connected correlation functions some of the leading
terms above are absent. For instance the connected four point function
has the expansion
\[
<\hat{A}_{4}(t)>_{c}\sim\hbar^{2}(b_{21}\hbar+b_{22}\hbar^{2}+\ldots)
\]
provided of course the $z^{4}$ term in $H_{I}$ is suppressed. On
the other hand the quartic term in $H_{I}$ may need to be retained
if though suppressed relative to the cubic term, it is competitive
with the one-loop ($b_{21}$) term. Also the two point function and
hence the power spectrum has the expansion,

\[
P\sim\hbar(b_{10}+b_{11}\hbar+\ldots).
\]
In particular this implies that 
\begin{eqnarray}
f_{NL}\sim\frac{<\hat{A}_{3}(t)>}{P^{2}(t)} & \sim & a'_{11}+\hbar a'_{12}+O(\hbar^{2})\label{eq:A3P2}\\
g_{NL}\sim\frac{<\hat{A}_{4}>_{c}}{P^{3}(t)} & \sim & b'_{21}+\hbar b_{22}^{'}+O(\hbar^{2})\label{eq:A4P2}
\end{eqnarray}
Let us now replace the ``quantum distribution'' (i.e. effectively
the classical distribution that is equivalent to the standard Bunch-Davies
vacuum distribution), by some classical distribution i.e. $\hbar\rightarrow{\cal A}$
($=l_{s}^{2}/\kappa^{2}$for instance) as discussed earlier. What
we see from the above is that while the leading (tree level) terms
are unaffected, the higher order (loop) effects are changed by factors
of ${\cal A}/\hbar$ which in the case of ${\cal A}=l_{s}^{2}/\kappa^{2}$
can be quite large for large volume string compactifications. Thus
in principle at least a classical distribution with a significantly
different unit of action (such as that coming from classical string
theory) can be distinguished from the ``quantum'' one with the unit
of action $\hbar$. However as we've argued, this by itself does not
test uniquely quantum features of quantum mechanics (as opposed to
classical distributions governed by the same unit of action). 

The above situation should be contrasted with that obtained with the
usual initial conditions but with $c_{s}\ll1$ and/or significant
dissipation during inflation $\gamma\gg H$. In this case it has been
estimated that \citep{LopezNacir:2011kk}
\begin{equation}
f_{NL}\equiv\frac{<\hat{A}_{3}(t)>}{P^{2}(t)}\sim\frac{\gamma}{c_{s}^{2}H}\gg1.\label{eq:fNLwarm}
\end{equation}
The current observational constraints on non-Gaussianities implies
that $|f_{NL}|\lesssim O(10)$. Thus very low sound speed and significant
dissipation appears to be ruled out. On the other hand replacing an
initial Gaussian distribution governed by $l_{P}$ by one that is
governed by $l_{s}$ (with $c_{s}\lesssim1,\,\gamma\lesssim H)$ will
not change the leading order contribution to $ $the bi- or the tri-spectrum.

\section{Conclusions}

In this paper we have argued that the usual quantum field theory calculation
of comological correlation functions is competely equivalent to calculating
these functions using a Gaussian statistical distribution governed
by the kernel \eqref{eq:Kstandard}. This may have been the result
of a stage prior to the onset of inflation such a state of quantum
(stringy) cosmology but knowledge of that is not relevant to the mathematics
of the derivation of inflationary fluctuations. We have then explored
the consequences of replacing $\hbar$ in the initial statistical
distribution by some other unit of action ${\cal A}$. In particular
we discussed the consequences of identifying ${\cal A}$ with a natural
unit of action coming from string theory and involving only classical
(but string theoretic) parameters. We noted how this could be consistent
with the usual double expansion of low energy string theory - namely
the $\alpha'$ expansion and the string loop expansion. One consequence
of this replacement is to change the relationship between the power
spectra and the height of the inflaton potential. We then discussed
in some detail alternative scenarios in which such a difference would
occur - namely situations with low speed of sound and/or a phase of
warm inflation. These were shown to be clearly different from the
type of change in the initial configuration that we have discussed
here.

We also noted that the difference between the standard prescription
for calculating the higher point functions and any other distribution
that is significantly different (i.e. with ${\cal A}/\hbar$ either
$\ll1$ or $\gg1$ as in the classical string theory case), will emerge
at higher orders in the loop corrections. The point is that what is
being tested in observations of the power spectrum and higher point
spectra, is a statistical distribution of decohered trajectories.
In other words there is no need at all to think of the initial state
for inflation as a pure quantum mechanical state. The entire discussion
of ``quantum fluctuations'' can be rephrased in terms of a decohered
initial state with a certain statistical weight. Whether or not that
weight corresponds to that arising from a pre-inflationary quantum
state of a simple QFT (which decohered before the onset of inflation),
rather than some distribution which is of stringy origin, cannot be
definitively established with current (or foreseeable future) measurements.
However distributions with values of ${\cal A}$ that are significantly
larger than $\hbar$ (such as one that may arise from a large volume
compactification of string theory), may possibly be ruled out by future
observations of non-Gaussianity.

\section{Acknowledgments}

I wish to thank Oliver DeWolfe, Salman Habib, Shamit Kachru, Fernando
Quevedo for discussions and especially Ramy Brustein, and Will Kinney
for discussions and comments on the manuscript and David Oaknin for
comments on the manuscript. Special thanks are also due to Paolo Creminelli
for comments on a previous version of this paper which led me to include
a discussion of higher point functions. Finally I wish to thank the
Abdus Salam ICTP for hospitality during the inception of this project.
This research was partially supported by the United States Department
of Energy under grant DE-FG02-91-ER-40672.

\section*{Appendix 1 Quantum and Classical Hamiltonian evolution}

Let us write the canonical dynamical variables $q^{i},p_{i},\,i=1,\ldots,n$
as 
\begin{eqnarray*}
\hat{z}^{i}(t) & = & \hat{q}^{i},\,i=1,\ldots,n\\
 & = & \hat{p}_{i-n},\,i=n+1,\ldots,2n.
\end{eqnarray*}
Here we've used hats to denote quantum operators satisfying the canonical
equal time commutation relations which in this notation read
\begin{equation}
[\hat{z}^{i}(t),\hat{z}^{j}(t)]=i\hbar J^{ij},\label{eq:CCR}
\end{equation}
where 
\[
\boldsymbol{J}=\begin{bmatrix}\boldsymbol{0} & \boldsymbol{I}\\
-\boldsymbol{I} & \boldsymbol{0}
\end{bmatrix}
\]
is the symplectic metric. The Heisenberg equations of motion are 
\begin{equation}
\frac{d\hat{z}^{i}(t)}{dt}=\frac{i}{\hbar}[H(\hat{z}(t),t),\hat{z}^{i}(t)].\label{eq:Heieqn}
\end{equation}
It is easily checked that the formal solution to this equation is
\begin{equation}
\hat{z}^{i}(t)=\hat{U}^{-1}(t,t_{0})\hat{z}^{i}(0)\hat{U}(t,t_{0}),\label{eq:hatzt}
\end{equation}
where 
\begin{equation}
\hat{U}(t,t_{o})=T\exp\left(-\frac{i}{\hbar}\int_{t_{0}}^{t}H(\hat{z}_{0},t')dt'\right)\label{eq:Uhat}
\end{equation}
with $T$ denoting time ordering. It is important to note that the
$\hat{z}$ in the Hamiltonian in this expression is evaluated at the
initial time $t_{0}$ ($\hat{z}_{0}\equiv\hat{z}(0)$. This is of
course only relevant because the Hamiltonians that we deal with have
explicit time-dependence. For any dynamical variable $\hat{A}(t)$
that is defined as a product of the canonical variables (with some
specified ordering if it involves both $q's$ and $p's$) there are
two alternate forms (see Weinberg \citep{Weinberg:2005vy}) for the
solution \eqref{eq:hatzt},
\begin{eqnarray}
\hat{A}(t) & = & \bar{T}e^{\frac{i}{\hbar}\int_{t_{0}}^{t}H(\hat{z}_{0},t')dt'}\hat{A}(t_{0})Te^{-\frac{i}{\hbar}\int_{t_{0}}^{t}H(\hat{z}_{0},t')dt'}\label{eq:orderedAHeis}\\
 & = & \sum_{N-0}^{\infty}\left(\frac{i}{\hbar}\right)^{N}\int_{t_{0}}^{t}dt_{1}\int_{t_{0}}^{t_{1}}\ldots\int_{t_{0}}^{t_{N-1}}dt_{N}\nonumber \\
 &  & [H(\hat{z}_{0},t_{N}),[H(\hat{z}_{0},t_{N-1})[\ldots[H(\hat{z}_{0},t_{1}),\hat{A}(t_{0})],\ldots]].\label{eq:commutAhatHeis}
\end{eqnarray}
Now let us separate the Hamiltonian into a quadratic (``free'')
part $H_{0}(t)$ and and an interaction part $H_{1}(t)$. The evolution
operator corresponding to the free Hamiltonian is 
\[
\hat{U}_{0}(t,t_{0})=T\exp\left(-\frac{i}{\hbar}\int_{t_{0}}^{t}H_{0}(\hat{z}_{0},t')dt'\right).
\]
 The interaction picture operators are defined by 
\begin{equation}
\hat{z}_{I}(t)=\hat{U}_{0}^{-1}(t,t_{0})\hat{z}_{0}\hat{U}_{0}(t,t_{0})\label{eq:zhatI}
\end{equation}
and the corresponding evolution operator defined by $\hat{U}_{I}\equiv\hat{U}_{0}^{-1}(t,t_{0})\hat{U}(t,t_{0})$
satisfies the equation of motion,
\[
i\hbar\frac{d\hat{U}_{I}}{dt}=H_{1}(\hat{z}_{I}(t),t)\hat{U_{I}}\equiv\hat{H}_{I}(t)\hat{U}_{I}.
\]
This has the formal solution 
\begin{equation}
\hat{U}_{I}(t,t_{0})=T\exp\left(-\frac{i}{\hbar}\int_{t_{0}}^{t}H_{I}(\hat{z}_{I}(t'),t')dt'\right)\label{eq:UI}
\end{equation}
and \eqref{eq:orderedAHeis}\eqref{eq:commutAhatHeis} are replaced
by (see for example \citep{Weinberg:2005vy}),
\begin{eqnarray}
\hat{A}(t) & = & \bar{T}e^{\frac{i}{\hbar}\int_{t_{0}}^{t}H_{I}(\hat{z}_{I}(t'),t')dt'}\hat{A}_{I}(t)Te^{-\frac{i}{\hbar}\int_{t_{0}}^{t}H_{I}(\hat{z}_{I}(t'),t')dt'}\nonumber \\
 & = & \sum_{N-0}^{\infty}\left(\frac{i}{\hbar}\right)^{N}\int_{t_{0}}^{t}dt_{1}\int_{t_{0}}^{t_{1}}\ldots\int_{t_{0}}^{t_{N-1}}dt_{N}\label{eq:orderedAI}\\
 &  & [H_{I}(\hat{z}_{I}(t_{N}),t_{N}),[H(\hat{z}_{I}(t_{N-1}),t_{N-1})[\ldots[H_{I}(\hat{z}_{I}(t_{1}),t_{1}),\hat{A}_{I}(t)],\ldots]].\label{eq:commutAI}
\end{eqnarray}
In the first equation above $T$ is time ordering while $\bar{T}$
is anti-time ordering. All this (except perhaps the second form of
the expressions for $\hat{A}$) is quite familiar. What may not be
so well known is that there is an exact classical analog of all these
equations. 

The classical variables $z^{i}$ satisfy the Poisson bracket relations
\begin{equation}
\{z^{i}(t),z^{j}(t)\}\equiv J^{ik}\frac{\partial z^{j}}{\partial z^{k}}=J^{ij}.\label{eq:zPB}
\end{equation}
Hamilton's equation of motion may then be written as,
\begin{equation}
\dot{z}^{i}(t)=\{z^{i}(t),H(z(t),t)\}=J^{ij}\partial_{j}H(z(t),t)=\partial^{i}H(z(t),t),\label{eq:Hameqn}
\end{equation}
where we've defined $\partial^{i}\equiv J^{ij}\partial_{i},\,\partial_{i}\equiv\partial/\partial z^{i}$.
This equation can be rewritten in the form of a commutator by introducing
the Hamiltonian vector field $H^{V}(z(t),t)\equiv\partial^{i}H(z(t),t)\partial_{i}$:
\begin{equation}
\dot{z}^{i}(t)=[H^{V}(z(t),t),z^{i}(t)]\label{eq:HVeqn}
\end{equation}
The solution to this equation is exactly the same as \eqref{eq:orderedAHeis}\eqref{eq:commutAhatHeis},
except that there are no factors of $i/\hbar$ and the operator Hamiltonian
is replaced by the vector field. In other words 
\begin{equation}
\frac{i}{\hbar}\hat{H}\rightarrow H^{V},\label{eq:HhatHV}
\end{equation}
and 
\begin{eqnarray}
A(t) & = & \bar{T}e^{\int_{t_{0}}^{t}H^{V}(z_{0},t')dt'}A(t_{0})Te^{-\int_{t_{0}}^{t}H^{V}(z_{0},t')dt'}\nonumber \\
 & = & \sum_{N-0}^{\infty}\int_{t_{0}}^{t}dt_{1}\int_{t_{0}}^{t_{1}}\ldots\int_{t_{0}}^{t_{N-1}}dt_{N}\nonumber \\
 &  & [H^{V}(z_{0},t_{N}),[H^{V}(z_{0},t_{N-1})[\ldots[H^{V}(z_{0},t_{1}),A(t_{0})],\ldots]].\label{eq:commutAclass}
\end{eqnarray}
 Of course all we've done here is to reverse the procedure of Dirac
who replaced the Poisson brackets of classical mechanics by $-\frac{i}{\hbar}$
times the commutator of the quantum operators. The point of the exercise
is simply to show that the evolution of a quantum operator represented
as an infinite series in terms of commutators, has an exact analog
in the classical theory. In fact up to operator ordering ambiguities
the relation between $\hat{A}(t)$ and $\hat{A}(0)$ is exactly the
same as that between their classical versions $A(t)$ and $A(0)$.
This is easily seen by comparing \eqref{eq:commutAhatHeis} and \eqref{eq:commutAI}.
Any commutator term in the first of these is of the form
\begin{equation}
\frac{i}{\hbar}[H(\hat{z}_{0},t'),\hat{z}_{0}^{j}]=\frac{\partial\hat{H}(\hat{z}_{0},t')}{\partial\hat{z}_{0}^{i}}\frac{i}{\hbar}[\hat{z}_{0}^{i},\hat{z}_{0}^{j}]=-\frac{\partial\hat{H}(\hat{z}_{0},t')}{\partial\hat{z}_{0}^{i}}\{z_{0}^{i},z_{0}^{j}(t)\}.\label{eq:HzQM}
\end{equation}
The first equality follows from the fact that the canonical equal
time commutator of two fields is a c-number while the second follows
from the Dirac identification between equal time commutators and (equal
time) Poisson brackets.

On the other hand the corresponding term in \eqref{eq:commutAI}is
\begin{equation}
[J^{ki}\frac{\partial H(z_{0},t')}{\partial z_{0}^{i}}\frac{\partial}{\partial z_{0}^{k}},z_{0}^{j}]=-\frac{\partial H(z_{0},t')}{\partial z_{0}^{i}}J^{ik}\frac{\partial z_{0}^{j}}{\partial z_{0}^{k}}=-\frac{\partial H(z_{0},t')}{\partial z_{0}^{i}}\{z_{0}^{i},z_{0}^{j}\}.\label{eq:Hzclass}
\end{equation}
So the two expressions \eqref{eq:HzQM}\eqref{eq:Hzclass} are the
same up to the replacement $\hat{z}\rightarrow z$ and so verifies
the statement above of the equality of the quantum and classical evolutions
up to operator ambiguities.

Clearly all the manipulations which led to the interaction picture
will survive with the replacement \eqref{eq:HhatHV}, so for the classically
evolved field we get exactly the same equations as \eqref{eq:orderedAI}\eqref{eq:commutAI}.
i.e.
\begin{eqnarray}
A(t) & = & \bar{T}e^{\int_{t_{0}}^{t}H_{I}^{V}(z_{I}(t'),t')dt'}A_{I}(t)Te^{-\int_{t_{0}}^{t}H_{I}(\hat{z}_{I}(t'),t')dt'},\label{eq:AIHVordered}\\
 & = & \sum_{N-0}^{\infty}\int_{t_{0}}^{t}dt_{1}\int_{t_{0}}^{t_{1}}\ldots\int_{t_{0}}^{t_{N-1}}dt_{N}\nonumber \\
 &  & [H_{I}^{V}(z_{I}(t_{N}),t_{N}),[H^{V}(z_{I}(t_{N-1}),t_{N-1})[\ldots[H_{I}^{V}(z_{I}(t_{1}),t_{1}),A_{I}(t)],\ldots]],\label{eq:AIHVcomm}
\end{eqnarray}
with
\begin{eqnarray}
z_{I}(t) & = & (U_{0}^{V})^{-1}(t,t_{0})z_{0}U_{0}^{V}(t,t_{0}),\label{eq:Iclassical}\\
U_{0}^{V}(t,t_{0}) & = & T\exp\left(-\int_{t_{0}}^{t}H_{0}^{V}(z_{0},t')dt'\right).\label{eq:U0classical}
\end{eqnarray}

These arguments are trivially extended to field theory. As usual in
the theory of cosmological fluctuations, one expands the original
generally covariant Lagrangian around the (time-dependent) inflationary
background and gets a time-dependent Hamiltonian functional of the
fluctuations. Denote the latter by
\begin{eqnarray}
z^{i}(\boldsymbol{x},t) & = & \phi^{i},\,i=1,\ldots,n\label{eq:fields}\\
 & = & \pi_{i-n},\,i=n+1,\ldots,2n,\label{eq:conjugates}
\end{eqnarray}
where $\phi,\pi$ are canonically conjugate field, field momentum.
The Hamiltonian vector field is 
\[
H^{V}(z(t),t)\equiv\int d^{3}xJ^{ij}\frac{\delta H[z(t),t)}{\delta z^{j}(\boldsymbol{x},t)}\frac{\delta}{\delta z^{i}(\boldsymbol{x},t)}.
\]
With this definition one can take over all the classical mechanics
formulae above to field theory just as the corresponding QM formulae
can be taken over to QFT. 

To proceed further we replace the expectation values of QFT with statistical
expectation values with some initial distribution $p(\phi_{0})$ .
i.e.
\begin{equation}
<\Omega|\hat{A}|\Omega>\rightarrow\int[d\phi_{0}]p(\phi_{0})A[\phi_{0},t)\label{eq:QMtoClass}
\end{equation}
where the second factor in the integrand on the RHS is to be calculated
using \eqref{eq:AIHVordered} or \eqref{eq:AIHVcomm}. The point is
that in both the left hand side and the right hand side of this relation
one evaluates in the interaction picture, using \eqref{eq:orderedAI}
or \eqref{eq:commutAI} for the LHS and \eqref{eq:AIHVordered} or
\eqref{eq:commutAclass} for the RHS. Thus one just has to calculate
expectation values of free fields and by Wick's theorem it is a sum
of products of two point functions determined by the correlator $<b_{\boldsymbol{k}}^{^{\dagger}}b_{\boldsymbol{q}}>$.
Hence given what we have just established on the time evolution of
classical and quantum operators, the LHS and the RHS of the above
relation are actually equal in value when $p(\phi)$ is defined as
in \eqref{eq:probhbar}, as discussed in section \eqref{sec:Comparison-with-cosmological}.

As discussed in section \eqref{sec:Comparison-with-cosmological}
with the distribution $p$ given by \eqref{eq:probhbar} we get exactly
the usual ``quantum'' calculation. What is evident from this discussion
is that all that is quantum here is the use of Planck's constant $\hbar$
in the expression for the kernel \eqref{eq:Kstandard} of the distribution.
One could equally well have started with a different initial distribution
as pointed out in that section. Also it is clear that the possible
operator ordering issues that might account for a difference in the
quantum and classical evolution will be irrelevant in a calculation
of the above expectation value at least if we start from an initial
Gaussian distribution since the interaction picture formulation in
either the quantum or the classical case, shows that the final result
is given by sums of products of Wick contractions. The difference
in the usual computations and one with some arbitrary classical distribution
is simply obtained by the replacement of the factor $\hbar$ in each
Wick contraction (i.e. two point function) by some other unit of action
${\cal A}$. \bibliographystyle{apsrev}
\bibliography{myrefs}

\end{document}